\documentstyle{mn} 
\input{epsf.tex}
\newcommand{\be}{\begin{equation}}                       
\newcommand{\ee}{\end{equation}}                        
\newcommand{\ba}{\begin{eqnarray}}                       
\newcommand{\ea}{\end{eqnarray}}                        
\newcommand{\brr}{\begin{array}}                        
\newcommand{\err}{\end{array}}                         
\newcommand{\bc}{\begin{center}}                        
\newcommand{\ec}{\end{center}}

\newcommand{\iras}{{\sl IRAS\/}}
\def\etal{et al.\ }
\def\kms{\ifmmode {\rm \ km \ s^{-1}}
\else
$\rm km \ s^{-1}$\fi}
\def\gtsima{$\; \buildrel > \over \sim \;$}
\def\ltsima{$\; \buildrel < \over \sim \;$}
                                       
\title[Two-point correlation function and morphological segregation in ORS]                    
{The two-point correlation function and morphological segregation in the {\em Optical Redshift Survey}}                         
\author[Hermit \etal]                        
{Sune Hermit,$^{1, 2}$ Bas\'{\i}lio X. 
Santiago,$^2$ Ofer Lahav,$^2$
Michael A. Strauss,$^{3,7}$ \\
\vspace{-1mm}\\
{\LARGE Marc Davis,$^4$ Alan Dressler$^{5}$ and John P. Huchra$^6$}\\
$^1$ Astronomical Observatory, Juliane Maries Vej 30, 2100 Kbh. 
{\O}, Denmark \\
$^2$ Institute of Astronomy,
Cambridge University, Madingley Road, Cambridge CB3 0HA, United
Kingdom\\
$^3$ Dept. of Astrophysical Sciences, Princeton University, Princeton, NJ 08544, U.S.A.\\
$^{4}$ Physics and Astronomy Departments, University of California, 
Berkeley, CA 94720, U.S.A.\\
$^{5}$ Observatories of the Carnegie Institution of Washington, 813 
Santa Barbara Street, Pasadena CA 91101, U.S.A.\\
$^{6}$ Center for Astrophysics, 60 Garden Street, Cambridge, MA 02138,
U.S.A.\\
$^{7}$ Alfred P. Sloan Foundation Fellow}                              
                                        
\date{July 26, 1996}

\begin{document}                                
                                        
\maketitle                                       
 
\begin{abstract}                                        
We study the clustering of galaxies in real and redshift space using 
the Optical Redshift Survey (ORS). We estimate the two point correlation 
function in redshift space, $\xi(s)$, for several subsamples of ORS, 
spanning nearly a factor of 30 in volume. We detect significant variations 
in $\xi(s)$ among the subsamples covering small volumes. For volumes 
\gtsima $(75 h^{-1} {\rm Mpc})^{3}$, however, the ORS subsamples present 
very similar clustering patterns. 
Fits of the form$\xi(s)=(\frac{s}{s_{0}})^{-\gamma_{s}}$ give best-fit 
values in the range $1.5 \leq \gamma_{s} \leq 1.7 $ and $6.5 \leq s_{0} 
\leq 8.8 h^{-1}$ Mpc for several samples extending to redshifts of 
8000 km s$^{-1}$. However, in several cases $\xi(s)$ is not well described 
by a single power-law, rendering the best-fit values quite sensitive to 
the interval in $s$ adopted. We find significant differences in clustering 
between the diameter-limited and magnitude-limited ORS samples within a 
radius of 4000 km s$^{-1}$ centered on the Local Group; $\xi(s)$ is larger 
for the magnitude-limited sample than for diameter-limited one. We 
interpret this as an indirect result of the morphological segregation
coupled with differences in morphological mix. We split ORS into 
different morphological subsamples and confirm the existence of 
morphological segregation of galaxies out to scales of $s \sim 10 h^{-1}$ 
Mpc. Our results indicate that the relative bias factor between early 
type galaxies and late-types may be weakly dependent on scale. If real, 
this would suggest non-linear biasing.

We also compute correlations as a function of radial and projected 
separations, $\xi(r_p, \pi)$, from which we derive the real space 
correlation function, $\xi(r)$. We obtain values $ 4.9 \leq r_{0} 
\leq 7.3 h^{-1}$ Mpc and $1.5 \leq \gamma_{r} \leq 1.7$ for various 
ORS samples. As before, these values depend strongly on the range in $r$ 
adopted for the fit. The results obtained in real space confirm those 
found using $\xi(s)$, i.e. in small volumes, magnitude limited samples 
show larger clustering than do diameter limited ones. There is no 
difference when large volumes are considered. Our results prove to be robust 
to adoption of different estimators of $\xi(s)$ and to alternative methods 
to compensate for sampling selection effects.
\end{abstract} 

\section{Introduction}
Early redshift surveys showed that 
galaxies are not distributed randomly in the Universe (Oort 1983). 
They are found to lie in clusters, filaments, bubbles and sheet-like 
structures (Davis \etal 1982, de Lapparent, Geller \& Huchra 1986,
Pellegrini \etal 1989). There exist large regions of the Universe which 
are almost devoid of galaxies (Kirshner \etal 1981, Geller \& Huchra 1988, de Lapparent, Geller \& Huchra 1988). 
This wealth of information about galaxy clustering and 
general distribution properties provide 
constraints on theories of galaxy and large-scale structure formation 
(Peebles 1980, Dekel \& Rees 1987, Strauss \& Willick 1995).

To characterize the distribution of galaxies and to 
quantify the degree to which it
departs from a Poisson distribution, a number of statistical tools have been 
developed. 
The first and most widely used approach to quantify the degree
of clustering in a galaxy sample has been
the two-point correlation function (Totsuji \& Kihara 1969, 
Peebles 1980). In spite of its known limitations, 
it has been applied to all redshift surveys completed
to date (e.g. Moore \etal 1994, 
Mart\'{\i}nez \& Coles (1994) [\iras\  QDOT], Dalton \etal 1994 
[APM Clusters], Loveday \etal 1995 [Stromlo-APM], de Lapparent \etal 1988, 
Park \etal 1994 [CfA2], Fisher \etal 1994 [\iras\  1.2 Jy]).
Similarly, higher-order clustering properties have been studied
by means of the three and four-point correlation functions or
by using counts in cells statistics (e.g. Efstathiou, 
Sutherland \& Maddox 1990, Bouchet \etal 1993). 
Recent efforts have also been invested into directly
reconstructing the galaxy density field (Saunders \etal 1991, Scharf \etal 1992, Strauss \etal 1992, Hudson 1993, Fisher \etal 1995b) 
or studying its power-spectrum (Vogeley \etal 1992, Fisher \etal 1993, da Costa \etal 1994). 

A limiting factor in comparing observed galaxy clustering properties
with model predictions is the difficulty in assessing whether
the observations reflect the 
properties of the ensemble of galaxies in the universe
(de Lapparent \etal 1988). 
The search for a `fair sample' of galaxies 
has led to the completion of ever larger
redshift surveys, probing both depth and angular extent.
A recent step towards this goal was the completion of
the Optical Redshift Survey (ORS), covering 98\% of the sky
for $\vert b \vert \geq 20^{o}$ and containing about 8500
optically selected 
galaxies (Santiago \etal 1995, 1996). ORS is
currently the closest 
approximation to an all-sky redshift survey selected in the optical 
and provides a dense sampling
of the galaxy distribution out to 8000 \kms.

The particular selection criteria used to define a redshift sample
may introduce systematic effects
on the derived clustering properties of galaxies.
Most samples 
are drawn from galaxy catalogues by applying 
either a diameter or a magnitude cut-off limit to them.
Thus, it is important to establish if diameter-limited and
magnitude-limited surveys sample the general galaxy population in a uniform 
way. In fact, comparisons between diameter and magnitude limited samples have
been made by Zucca \etal (1991) but with much smaller samples.
ORS is again an excellent dataset for such analyses, given that it
is essentially complete in both diameters ($ \theta > 1\farcm 9$) and magnitudes
($m_B \leq 14.5$).

Finally, an important 
observational result of previous galaxy clustering analyses
is the existence of segregation as a function of morphological type.
It has been qualitatively known for decades
that elliptical or bulge-dominated (`early-type') galaxies prevail in the
cores of rich clusters of galaxies whereas disk-dominated galaxies 
(spirals) make up most of the general field population (Hubble \& Humason 1931,
Abell 1965). This result
is often described as the so called
morphology-density relation, first quantified by Dressler
(1980) for clusters, 
and also known to apply for smaller galaxy concentrations such
as groups (Postman \& Geller 1984, Ferguson \& Sandage 1991).
A morphology-radius
relation have also been suggested (Whitmore, Gilmore \& Jones 1993).
Indications that morphological segregation also exists among field galaxies
have been given by several
works (Davis \& Geller 1976, Giovanelli, Haynes \& Chincarini 1986,
Eder \etal 1989, Lahav, Nemiroff \& Piran 1990,  Mo \& B{\"o}rner 1990, 
Santiago \& Strauss 1992).
While morphological segregation is fairly well 
established, at least on scales up to $\sim 10 h^{-1}$ Mpc
($H_{0}= 100 h$ km s$^{-1}$ Mpc$^{-1}$), 
it has been more difficult to establish if the 
clustering properties of galaxies depend on luminosity 
(Phillipps \& Shanks 1987, Alimi, Valls-Gabaud \& Blanchard 
1988, Davis \etal 1988, Hamilton 1988, Salzer, Hanson \& Gavazzi 1990, 
Iovino \etal 1993, Park \etal 1994), 
surface brightness (Davis \& Djorgovski 1985, Santiago \& da Costa 1990,
Thuan \etal 1991), spectral properties (Salzer \etal 1988)
or circular velocity (White, Tully \& Davis 1988, Mo \& Lahav 1993). 
On the theoretical side, the existence of segregation among galaxies
and the indirect evidence for segregation between light and matter on 
galactic and extra-galactic scales has led to the formulation
of scenarios of biased galaxy formation (Kaiser 1984, Babul \& White
1991).

In this paper we address the issues of clustering dependence on survey
volume, sample selection criteria and morphology.
The paper is structured as follows. In section \ref{sec.Data} we 
briefly introduce the ORS data. In section \ref{sec.Two} the 
different estimators of the two-point correlation function and their
associated errors are 
presented. We also discuss different schemes to compensate for selection
effects inherent to the data. 
In section \ref{sec.compare} we present results based on the comparison of
the correlation functions in redshift space, $\xi(s)$, 
for different ORS subsamples split according to volume, selection
criteria and morphological types.
In section \ref{sec.xir} we compute $\xi(r_p,\pi)$, the 
correlation as a function of radial and projected separations, 
and invert it to obtain the real-space correlation, $\xi(r)$. 
Our results for $\xi(r)$ are consistent with those of section
\ref{sec.compare}. Finally, in section \ref{sec.Conclusions} we 
state our main  
conclusions and discuss our results.

\begin{figure*}
\epsfxsize=17cm
\epsfbox{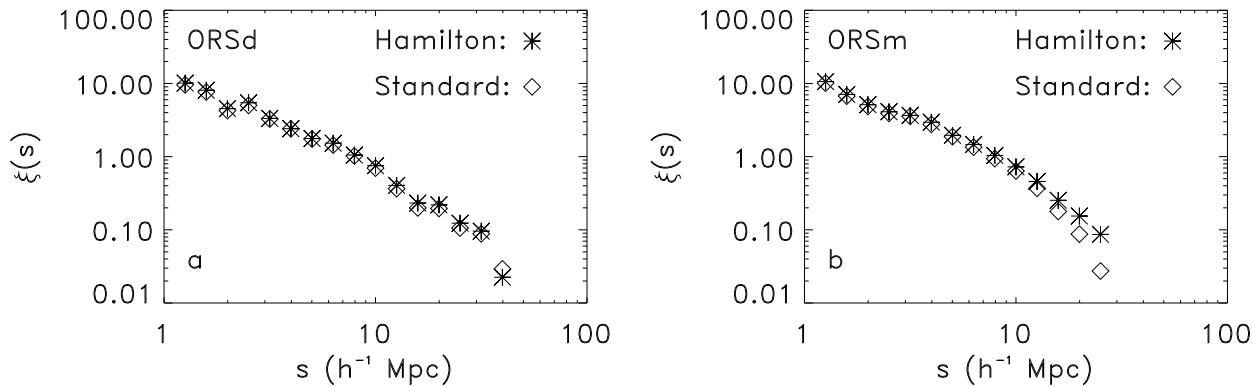}
\caption[]{The correlation function for ORSd (panel a) and 
ORSm (panel b) out to $V_{max} = 8000 \kms$
using the Hamilton (stars) and Standard 
estimator (diamonds).}
\label{fig.ORSdHS1.ps}
\end{figure*}

\section{The Data}\label{sec.Data}
ORS was presented by Santiago \etal 1995 (hereafter Paper I).
In brief, the sample was drawn from three catalogues: the Uppsala 
General Catalogue of Galaxies (UGC; Nilson 1973), 
the European Southern Observatory Galaxy Catalogue (ESO; 
Lauberts 1982, Lauberts \& Valentijn 1989) and the 
Extension to the Southern Galaxy Catalogue (ESGC; Corwin \& Skiff 1994). 
ORS contains two largely overlapping subsamples, one 
magnitude limited to $m_B \leq 14.5$
(hereafter ORSm) and the other diameter limited
to $\theta_B \geq 1.9'$ (hereafter ORSd). 
ORSd was created by merging ESGC, ESOd and UGCd, and ORSm 
consists of ESOm and UGCm. The entire ORS catalog consists 
of 8457 objects; redshifts are available for 8266.
To complete the sample, roughly 1300 new redshifts were 
measured -- the remaining redshifts were taken from the 
literature. Most of the new redshifts were obtained in regions of
low galactic latitude, $20^{\circ} \leq |b| \leq 30^{\circ}$, 
and within the strip not covered by either the UGC or ESO catalogues
($-17$\fdg 5${}\leq \delta \leq-2$\fdg 5). ORS provides the most 
detailed and homogeneous sampling to date of these areas of 
the sky. The matching of the catalogs and the derivation of 
the luminosity and diameter functions for the ORS are treated 
in Santiago \etal (1996) (hereafter Paper II).
 
\section{The Two-point Correlation Function}\label{sec.Two}
Consider a Poissonian distribution of points in space. The 
joint probability that one point is found in each of 
two independent volume elements $dV_{1}$ and $dV_{2}$ obviously scales 
with the mean density and the size of the volume elements
\be
dP=\bar{n}^{2}dV_{1}dV_{2}.
\ee
If the distribution is non-Poissonian then there may be an 
excess probability of finding a point in each of the two volume 
elements $dV_{1}$ and $dV_{2}$ separated by the distance $r$ 
(see, e.g Peebles 1980):
\begin{equation}
dP=\bar{n}^{2}dV_{1}dV_{2}[1+\xi(r)].
\label{eq.dp}
\end{equation}
This excess probability, quantified by $\xi(r)$,
is the two-point correlation function. 

Traditionally, $\xi(r)$ has been determined by the expression
\begin{equation}
\xi_{S}(r)=\frac{\bar{n}_{R}}{\bar{n}_{D}} \frac{DD(r)}
{DR(r)} - 1,
\end{equation}
where $DD(r)$ and $DR(r)$ are, respectively, the
number of galaxy-galaxy and galaxy-random pairs with separation 
within the range $r$ and $r+dr$. The latter are computed by creating a
catalogue of randomly distributed points within the same volume occupied
by the real galaxies.
$\bar{n}_{D}$ ($\bar{n}_{R}$) is the number 
density of galaxies (random points) in this volume.
We hereafter refer to $\xi_S (r)$ as the standard
estimator. In spite of its widespread usage, the standard estimator
explicitly depends on the mean density assigned to the sample. 
As the lack of a fair sample may bias $\bar{n}_D$, $\xi_S (r)$ may be 
(under) overestimated accordingly (de Lapparent \etal 1988, Davis \etal
1988). More recently, Hamilton (1993)
has proposed an alternative estimator, which is much less sensitive to 
this problem:

\begin{equation}
\xi_{H}(r) = \frac{DD(r) RR(r)}{[DR(r)]^{2}} -1,
\end{equation}
where $RR(r)$ is the random-random pair counts within the bin of 
separation $r$.
Hamilton (1993) shows that the dependence of $\xi_{H}(r)$ on
the mean density of 
galaxies in the survey volume is of second order, whereas the dependence
of $\xi_{S}(r)$ is of first order.
Thus $\xi_{H}(r)$ is less affected by the 
uncertainty in the mean density.

\begin{figure*}
\epsfxsize=17cm
\epsfbox{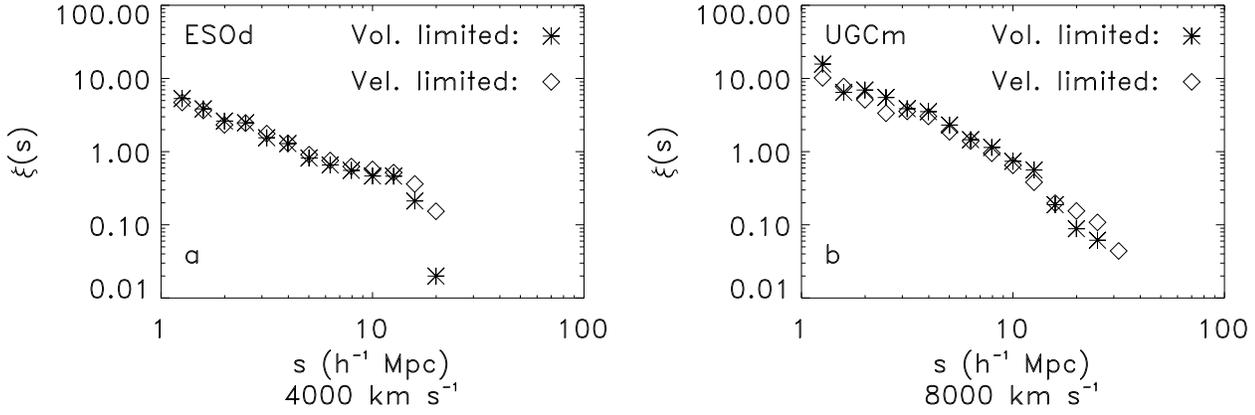}
\caption[]{Comparison of volume (stars) and velocity (diamonds) 
limited subsamples for ESOd (panel a) and UGCm (panel b). }
\label{fig.velvol}
\end{figure*} 

In Fig.~\ref{fig.ORSdHS1.ps}, we compare the two estimators in redshift
space, 
$\xi_{H}(s)$ and $\xi_{S}(s)$, for both ORSd and ORSm cut at
$V_{max} = 8000 \kms$. 
Throughout this paper, we adopt the usual convention and use
$\xi(s)$ ($\xi(r)$) to denote the correlation function in 
redshift (real) space.
The two estimators give very similar
results in both cases except on the largest scales, where 
the clustering is weak. A similar trend was found by Loveday 
\etal (1995), who analysed the Stromlo-APM survey. 
Since $\xi_{H}(s)$ gives a more reliable estimate of the 
two-point correlation function on all scales, 
we chose to use that estimator exclusively in what follows.

In practice, the pair counts $DD$, $DR$ and $RR$ are weighted sums
of pairs, which take into account the selection effects inherent to
the galaxy sample used (see \S 3.1 below). Likewise for
the mean number density $\bar{n}_D$. The catalogue of random points 
may or may not imitate such selection effects,
provided that the appropriate selection function is applied to
compensate for them. In this paper we did not introduce any
selection effects to the random points. The weighting scheme adopted
in this work is the topic of the next subsection.

\subsection{Weights and selection functions}
In a diameter or magnitude limited survey, 
the sample will become sparser at larger redshifts due to the 
increasing loss of galaxies caused by the adopted apparent 
magnitude or diameter cut-off. 
This selection effect has to be correctly accounted for when determining
$\xi(s)$ or any other statistic from the sample. 
The effect is quantified by the selection function, $\phi({\bf r})$, which 
expresses the fraction of the total population of galaxies 
that are expected to satisfy the sample's selection criterion
at any given point in space. Under the assumption
of a universal luminosity (or diameter) function, 
the selection function depends solely on distance.
For ORS, however, an angular dependence is 
introduced by variable (and non-negligible) amounts of Galactic extinction
as well as by non-uniformities in the magnitude or diameter systems among
the three different catalogues from which the data were derived. 
Paper II presents a detailed discussion of the problem and of the way 
these additional
selection effects were incorporated into $\phi({\bf r})$. In brief, 
extinction effects were incorporated into the maximum
likelihood approach used to fit $\phi({\bf r})$ (see also Yahil \etal 1991).
A selection function was also fitted separately 
to each of the five ORS subsamples (ESOm, ESOd, UGCm, UGCd and ESGC)
in order to bypass the inhomogeneity in selection caused by the different
catalogues.
 
Once $\phi({\bf r})$ is available, selection effects are in principle
compensated for by weighting each galaxy $i$ by $w_i = 1 / \phi({\bf r}_i)$.
However, the different subsamples in ORS have different diameter and/or
magnitude systems, which introduces an additional selection bias: subsamples
based on deeper plate material tend to contain more galaxies
and thus have a larger mean number density (Paper II). 
We thus normalize each subsample by the mean number density of
galaxies in that subsample. This would be adequate if each subsample
in itself were a fair sample, but that it not necessarily the case, 
and thus, as explained
in Paper II, we renormalize again by the mean density of each
subsample volume relative to the full sky, as found in the full-sky
\iras\ 1.2 Jy sample (Fisher \etal 1995a). 
We thus assign galaxy $i$ a weight given by
\begin{equation}	
w_{i} = \frac{1}{\phi({\bf r}_i) n_{k,ORS}} ~
\frac{n_{k,IRAS}}{n_{tot,IRAS}},
\end{equation}
where $n_{k,ORS}$ 
($n_{k,IRAS}$) is the mean number density
in the region corresponding to the $k^{th}$ subsample 
in ORS (IRAS) and $n_{tot,IRAS}$ 
is the total mean number density
as computed from the \iras\  1.2 Jy sample. All number densities correspond
to a volume-limiting distance of 500 km s$^{-1}$ (cf., Paper II). 

The weighting scheme used here is not minimum variance, in the sense of Davis
\& Huchra (1982) and Efstathiou (1988). However, minimum weighting 
requires prior knowledge of the correlation 
function. This problem has been bypassed in the past by usage of 
earlier estimates of $\xi(r)$ or of N-body simulations
(Loveday \etal 1992,1995).
For a dense sample like ORS, however, the
practical improvement provided by using a minimum
variance method is very small (Hamilton 1993, Park \etal 1994). 

Finally, we should point out that the results presented in the next sections
have been tested with different weighting schemes and prove to
be robust.

\subsection{Error estimation}
There is no standard method of calculating 
errors for correlation functions (Hamilton 1993, Strauss \& Willick 1995).
One could simply assume Poisson fluctuations in the pair counts, but since 
they are correlated on any given scale, this is not strictly correct. 
Instead we use 10 
bootstrap resamplings of the data (Barrow, Bhavsar \& Sonoda 
1984, Ling, Frenk \& Barrow 1986, Mo, Jing \& B{\"o}rner 1992) 
to estimate the variance in the calculated correlation functions. 
As shown by Fisher \etal (1994), this is likely to be an overestimate 
of the real errors and is thus appropriate given 
the conservative approach we adopt throughout this paper. 

\section{Clustering in ORS}\label{sec.compare}
The ORS subsamples used here are limited at $V_{max} = 8000 \kms$
in order to minimize shot noise. The selection function parameters
for those subsamples with $V_{max} = 8000 \kms$ are 
listed in Table 1 of Paper II.
As mentioned before, to calculate $\xi(s)$ we created
random samples with the same geometry as the data sample in 
consideration; in the case of ORSd, for example, we limited the random 
sample to 8000 km s$^{-1}$, $|b| \geq 
20^{\circ}$ and to regions with $A_B \leq 0.7$ as
determined from the extinction maps of Burstein \& Heiles (1982). 
When dealing with ESOm (and therefore ORSm) we also
took into account the lack of ESO-LV magnitudes in several
ESO plates and thus did not place random points 
within the boundaries of these plates (see Paper I for details).

We assigned unit weight to the random points and used
logarithmic binning on all scales. We tested the 
code and our random number generator 
by computing $\xi(s)$ on several pairs of random samples consisting of 
20000 points each. 
As expected, this yielded no correlations on scales $s \geq 0.5 h^{-1}$ Mpc. 
As a final test of our code, we calculated $\xi(s)$ for the 
\iras\  1.2 Jy sample (Fisher \etal 1995a) and successfully recovered the
results of Fisher \etal (1994).
 
\begin{figure}
\epsfxsize=8.3cm
\epsfbox{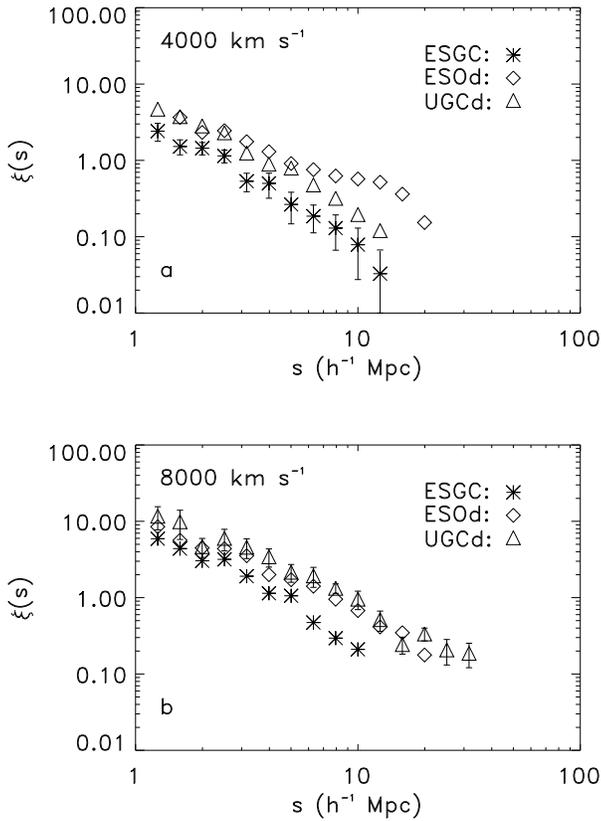}
\caption[]{Panel a: $\xi(s)$ for ESGC (stars), ESOd (diamonds) 
and UGCd (triangles) cut at $V_{max}= 4000$ km s$^{-1}$. 
Panel b: The same, but for subsamples cut
at $V_{max} = 8000$ km s$^{-1}$.}
\label{Fig4}
\end{figure}

\begin{figure}
\epsfxsize=8.3cm
\epsfbox{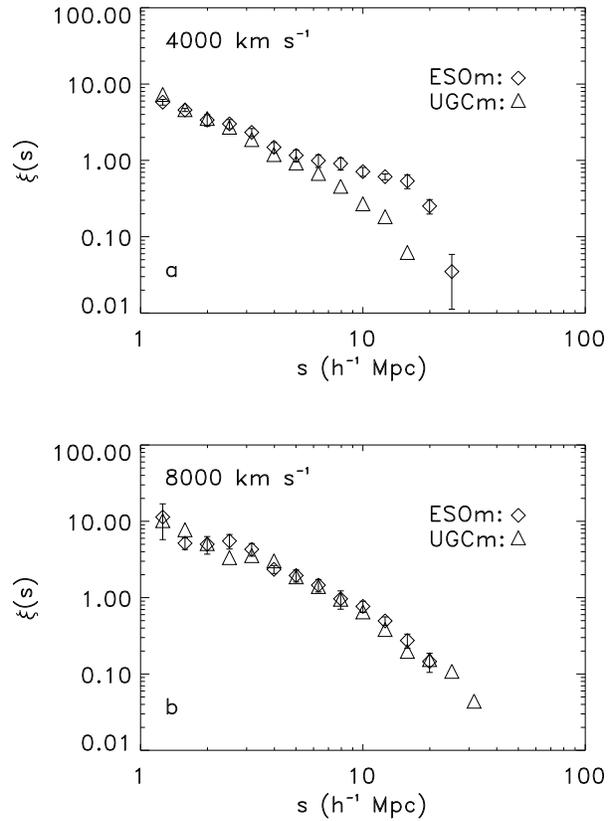}
\caption[]{The same as Fig.~\ref{Fig4} but for magnitude 
limited subsamples.}
\label{Fig5}
\end{figure}

An important issue is to test the reliability of the weights
assigned to each galaxy. We do that by comparing subsamples
cut at a given maximum distance $V_{max}$ (hereafter referred to as
velocity-limited at $V_{max}$) to their
volume-limited counterparts. These latter subsamples, by 
definition, include only objects which are luminous (or large) 
enough to satisfy the selection criteria of the
survey if they were at $V_{max}$. This is equivalent to applying a
lower limit to the luminosity (or absolute diameter) to the 
subsample, leading to a uniformly
selected dataset with unit weight assigned to each object
($\phi (r) = 1$, $0 \leq r \leq V_{max}$). Since both velocity and 
volume-limited subsamples probe the same volume in space, 
they should lead to very similar $\xi (s)$,
unless the weights in the velocity-limited subsample are biased 
or there is segregation as a function of luminosity (or absolute 
diameter). The results are plotted in Fig.~\ref{fig.velvol}. Panel
{\it a} shows ESOd out to  
$V_{max} = 4000$ km s$^{-1}$ 
and panel {\it b} refers to UGCm out to $V_{max} = 8000$ km 
s$^{-1}$.
The agreement is very good in both panels, especially when one 
considers the relatively small number of galaxies in the volume 
limited samples (386 and 433 objects in panels {\it a} and {\it b}, 
respectively). We are therefore confident that the 
selection function derived in Paper II is reliable
and hereafter continue working
with velocity-limited subsamples.

Finally, we note that redshifts 
are a combination of the
Hubble flow and peculiar velocities, the latter distorting 
the determination of galaxy clustering in redshift space.
In the remainder of this section we present correlation functions for
ORS in redshift space, $\xi(s)$.
Since the main purpose of this
work is to compare $\xi(s)$ for different samples, we adopt the
working hypothesis that, to first order,
peculiar velocities will affect the samples being compared in similar ways.
In section \ref{sec.xir} we calculate the real-space correlation
function, and find similar results to those in this section, confirming our
working hypothesis. 
 
\subsection{$\xi(s)$ for the different subsamples}
If a survey sampling the local universe contains structures whose sizes
are comparable to the volume of the survey itself, it is not considered a 
`fair sample' of the large-scale galaxy distribution. A detailed
discussion of the effect of such survey-sized structures is presented
by de Lapparent \etal (1988). 
Therefore, it is important to define the minimum survey volume which would
satisfy the condition for being a fair sample. 
Since ORS provides us with a dense sampling of the local universe over
two-thirds of the celestial sphere, we here investigate whether it
may be considered a fair sample, at least for the purpose of estimating
$\xi (s)$. We thus compute
$\xi (s)$ separately for the 5 ORS subsamples (ESOd, UGCd, ESGC, ESOm and
UGCm) both out to $V_{max} = 4000$ km s$^{-1}$ and 
$V_{max} = 8000$ km s$^{-1}$, 
with volumes spanning nearly a factor of 30. 

The results for the diameter-limited data 
are plotted in Fig.~\ref{Fig4} and the corresponding plots
for the magnitude limited subsamples are in 
Fig.~\ref{Fig5}. Errorbars denote one standard deviation
from the distribution of $\xi (s)$ for the 10 bootstrap 
resamplings mentioned in section 3.2.
For clarity, we have only plotted errors for one of the 
subsamples in each of the two panels in Fig.~\ref{Fig4} 
and Fig.~\ref{Fig5}. In panel \ref{Fig4}{\it a} we plot
errors for ESGC. It is the smallest of the ORS subsamples
and correspondingly the one that has largest associated errors.

The differences in $\xi (s)$ are significant for the
subsamples shown in panels \ref{Fig4}{\it a} and \ref{Fig5}{\it a}.
There are large
discrepancies between UGC and ESO on scales $ 7 - 
20 h^{-1}$ Mpc regardless of
selection criteria. 
ESGC seems to exhibit a slightly steeper slope and a 
much smaller correlation length than both the ESOd and UGCd samples.
For $V_{max} = 8000$ km s$^{-1}$, however, $\xi(s)$ is
very similar for the ESO and UGC subsamples. This is true
in both figures \ref{Fig4} and \ref{Fig5}.
The ESGC $\xi(s)$ still has a smaller amplitude at $V_{max} = 8000 \kms$ 
than do ESOd and UGCd.
ESGC, however, covers a volume less than 
half of that of ESOd and less than a third of UGCd.
We thus conclude that the ESOd and UGCd subsamples stretching out to
$V_{max} = 8000 \kms$ are close approximations to a fair sample 
(at least for the purpose of estimating $\xi(s)$ within $20 h^{-1}$ Mpc), 
whereas ESGC and the subsamples limited at $V_{max} = 4000 \kms$ are not.
The minimum volume that supposedly contains 
a representative sample of the ensemble, for these purposes, is thus 
$\sim (75 h^{-1} {\rm Mpc})^{3}$.
Notice that in both ESO and UGC, diameter and magnitude-limited subsamples
have a similar $\xi(s)$ slope.
The latter, however, are more clustered for $V_{max} = 4000 \kms$
but not for $V_{max} = 8000 \kms$. This issue will be revisited in
\S 4.2.

For each of the subsamples large enough to be considered good approximations
to a fair sample we fit a power-law of the form
\begin{equation}
\xi(s)=\left (\frac{s}{s_{0}} \right )^{-\gamma_s}
\end{equation}
and determine the best-fit value of the slope $\gamma_s$ and 
correlation length $s_{0}$ using a least squares fit.
The scale ranges used in the fits and the derived values are 
listed in Table \ref{tab.corr}. The best-fit values span the range 
$1.5 \leq \gamma_s \leq 1.7$ and $6.5 \leq s_0 \leq 8.8 h^{-1}$ Mpc. 
However, some of the cases shown in Fig.~\ref{Fig4} 
and Fig.~\ref{Fig5} are not
well described by a single power-law over all scales. The presence
of curvature in $\xi(s)$ is also visible
in the APM (Loveday \etal 1995) and quite pronounced in the
QDOT sample (Moore \etal 1994). Moore \etal (1994) hypothesize that 
this is caused by a combination of peculiar velocities and observational 
errors which will enhance the correlations on small scales and suppress 
them on larger scales.
Since the $\xi(s)$ are not perfect power-laws, the values listed in
Table \ref{tab.corr} are very 
sensitive to the scale interval used in the fits.
For instance, if we fit a power-law to $\xi(s)$ for ORSd (ORSm) 
limited at $V_{max} = 8000$ \kms over the 
range $1 \leq s \leq 10 h^{-1}$ Mpc, we find $s_{0}=8.19 (8.58)$ and 
$\gamma_{s}=-1.26 (-1.18)$. If the fitted range is 
$3 \leq s \leq 16 h^{-1}$ Mpc, however, the results are 
$s_{0}=7.36 (7.65)$ and $\gamma_{s}=-1.58 (-1.62)$.
The ranges listed in Table~\ref{tab.corr} are thus the ones where 
$\xi(s)$ was best fit by a single power-law. 
Also, due to the fact that the points are not statistically independent, 
it is not strictly correct to use a least square method for 
this purpose. However, as demonstrated by Bouchet \etal (1993), 
it is a good first approximation, and in any case the 
uncertainty in the fitted values will be dominated by the strong 
dependence on the range in $s$ chosen for the fit.  

\begin{figure}
\epsfxsize=8.3cm
\epsfbox{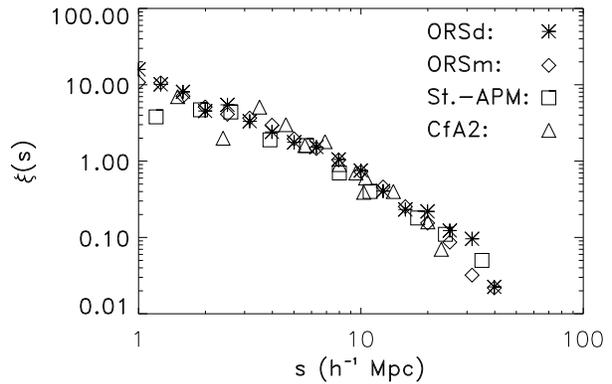}
\caption[]{Comparison of $\xi(s)$ for ORSd (stars) and 
ORSm (diamonds) out to $V_{max} = 8000 \kms$ with previous
estimates based on optical samples: Stromlo-APM (squares) and
CfA2 (triangles).}
\label{Figcomp}
\end{figure}

We now compare our $\xi(s)$ determinations with those of other authors.
We use the ORSd and ORSm correlations for that purpose. Power-law
fits to these latter are listed in the last two lines of Table \ref{tab.corr}.
Our $\xi(s)$ for ORSd and ORSm are 
in excellent agreement with the results obtained by 
de Lapparent \etal (1988) and by Loveday \etal
(1995) for the CfA2 and Stromlo-APM samples, respectively, 
as shown in Fig \ref{Figcomp}.
These previous works probe larger volumes than does ORS but contain a smaller
number of galaxies, hence our smoother $\xi(s)$.
Notice that the best-fit values to Eq. (5) quoted by Loveday \etal (1995)
are significantly different from those listed in Table \ref{tab.corr}.
This is entirely due to differences in the scale range used in the fits,
which illustrates the sensitivity of the fitted values on the interval in
$s$ used.
Finally, our ORSd and ORSm $\xi(s)$ are also in excellent agreement with
those determined by Davis \& Peebles (1983) and Maurogordato, Schaeffer \& da Costa (1992) for the CfA1 and SSRS samples, respectively.
This is not very surprising given the fact that the UGCm and CfA1 overlap
considerably, as do ESOd and SSRS.

\begin{table*}
\caption{Power-law fits and parameters}
\begin{tabular}{|l|c|c|c|c|c|c|} \hline
Sample & Limit  & Range  & $\gamma_{s}$& $s_{0}$ & $\gamma_{r}$ & $r_{0}$ \\
 & ($h^{-1}$ Mpc) &($h^{-1}$ Mpc) & & & & \\ 
ESOd    & 80 & 2.5 - 20  & 1.48    & 7.2 &- &- \\
ESOm    & 80 & 2.5 - 20  & 1.65    & 7.5 & -& - \\ 
UGCd    & 80 & 2.5 - 32 & 1.48    & 8.8  & -&- \\
UGCm    & 80 &1.6 - 32    & 1.64    & 6.5 & -&- \\
ORSd    & 80 & 1 - 40 & 1.57    & 6.6 & 1.59 & 7.07 \\
ORSm    & 80 & 1.3 - 25   & 1.51    & 6.8 & 1.56 & 7.26\\ \hline
\end{tabular} \label{tab.corr}
\end{table*}

\subsection{ORSd and ORSm}
\begin{figure}
\epsfxsize=8.3cm
\epsfbox{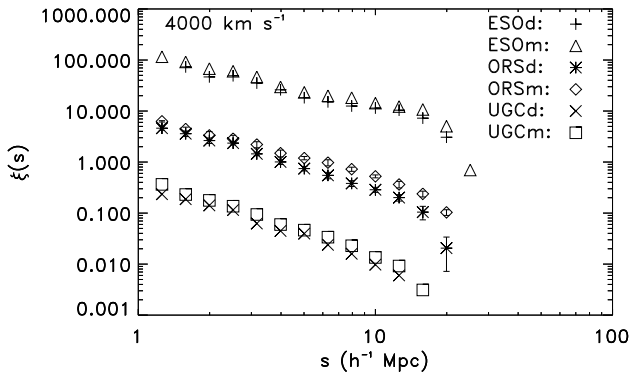}
\caption[]{The correlation function for ORSd (stars) and ORSm (diamonds) 
velocity limited at 4000 km s$^{-1}$. For comparison we also show
$\xi(s)$ for ESOd (plus signs), ESOm (triangles), UGCd (crosses) and 
UGCm (squares). These latter are offset by a factor of 20 for the sake
of clarity.}
\label{Fig6}
\end{figure}

\begin{figure}
\epsfxsize=8.3cm
\epsfbox{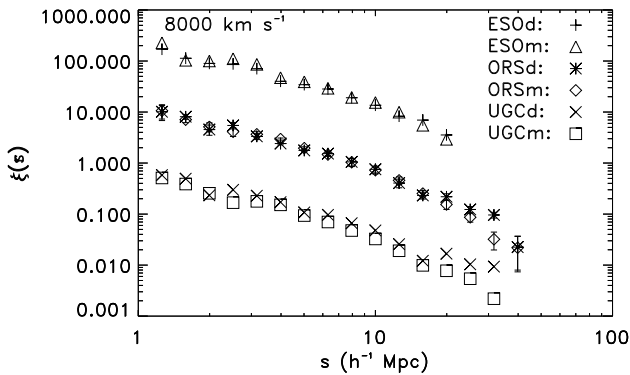}
\caption[]{The same as Fig.~\ref{Fig6}, only for samples velocity 
limited at 8000 km s$^{-1}$.}
\label{Fig7}
\end{figure}

The correlation function $\xi(s)$ for both ORSd and ORSm 
samples can be seen in Fig.~\ref{Fig6} and Fig.~\ref{Fig7} for
$V_{max} = 4000 \kms$ and $V_{max} = 8000 \kms$, respectively.
Also shown are the equivalent correlations for
ESO and UGC subsamples. These latter have been offset by
a constant for clarity.
The errorbars again indicate 1 $\sigma$ deviations as determined from the 10 
bootstrap resamplings. 
The ORSd and ORSm samples within 
$V_{max} = 4000$ km s$^{-1}$ have significantly 
different correlation functions.  
ORSm is more clustered than ORSd over the entire
range of scales available.
Similar behaviour is seen for both ESO and UGC; 
the magnitude limited subsamples are more clustered 
than their diameter-limited
counterparts in Fig.~\ref{Fig6}.
Thus, it appears that 
the difference in clustering amplitude between ORSd and ORSm arises 
from similar contributions from the ESO and UGC subsamples. 
On the other hand, for $V_{max} = 8000$ km s$^{-1}$, 
the ORSd and ORSm have very similar $\xi(s)$,
except on the very largest scales, where ORSm falls off 
faster than ORSd (Fig.~\ref{Fig7}).

The only substantial difference between the ORSm and ORSd volumes is the ESGC
strip ($-17.5^\circ < \delta < -2.5^\circ$), which is not included in
ORSm. Thus it is unlikely that the difference in the ORSm and ORSd
correlations to $4000 \kms$ is due to large-scale
differences in the underlying density field. To make sure, 
we removed ESGC from ORSd and computed
$\xi(s)$ again. Fig.~\ref{Fig6} remained essentially unchanged;
the vast majority of DD, DR and RR pairs in fact come from within ESOd, UGCd
or from ESOd/UGCd cross-boundary pairs.

\begin{figure}
\epsfxsize=8.3cm
\epsfbox{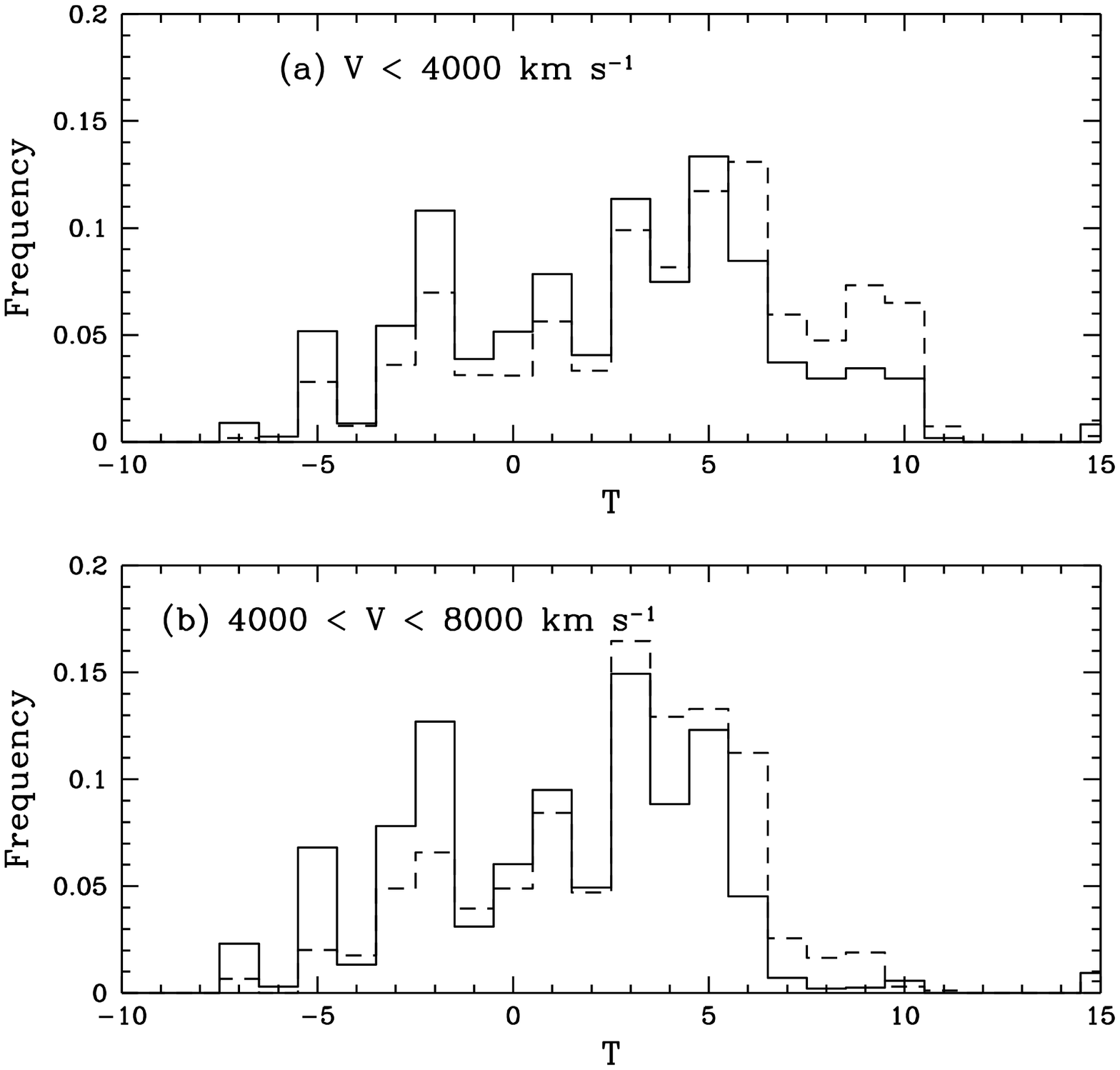}
\caption[]{The frequency of morphological type for ORSd (dashed lines) and ORSm (solid lines) at $V_{max} = 4000$ \kms (panel a) and $4000 < V_{max} < 8000$ \kms (panel b). The difference in morphological mix is clearly diminished at 
larger redshifts, rendering more similar results for $\xi(s)$.}
\label{fig.thist}
\end{figure}

The observed discrepancy between ORSm and ORSd in Fig.~\ref{Fig6} is
likely to reflect real clustering differences
between magnitude-limited and diameter-limited samples.
Since the former tend to pick up, on average, higher surface
brightness, earlier type galaxies than the latter, 
one expects ORSm to
show enhanced clustering as a result of the larger
fraction of stronger clustered early-type galaxies it contains.
The distributions of morphological types 
in ORSd (dashed lines) and ORSm (solid lines)
are shown in Fig.~\ref{fig.thist} 
for $V_{max} = 4000$ \kms 
(panel a) and $4000 < V_{max} < 8000$ \kms (panel b).
As expected, the morphological composition of 
ORSm is clearly skewed towards early-types relative to ORSd.
Thus, the larger clustering in ORSm seen in Fig.~\ref{Fig6} 
reflects segregation as a function of morphology and/or surface brightness.
As attested by panel {\it \ref{fig.thist} b},
the fraction of late-type, low surface brightness galaxies in ORSm and ORSd
is markedly reduced beyond $V \sim 4000$ \kms. This is expected to occur in
any flux-limited (or diameter-limited) sample since these galaxies tend also
to be less luminous (and smaller) than typical spirals and ellipticals.
The reduced difference in the morphological mix of ORSm and ORSd
for $V_{max} = 8000$ \kms renders their $\xi(s)$ very similar in this larger
volume.

\begin{figure}
\epsfxsize=8.3cm
\epsfbox{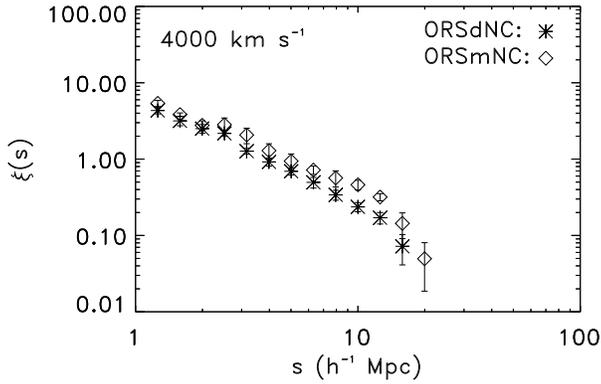}
\caption[]{The correlation function for the cluster-removed ORSd and 
ORSm samples velocity limited at $V_{max} = 4000$ km s$^{-1}$.}
\label{Fig8}
\end{figure}

Since morphological segregation is more pronounced in clusters,
we test the effect described above by removing
the 6 clusters shown in Table \ref{Tab.NC} and rederiving $\xi(s)$ for 
both ORSm and ORSd out to $V_{max} = 4000$ \kms.
This list contains the most prominent clusters
within this volume.
Columns 1, 2, 3 and 4 list the name and 
position of each cluster.
Columns 5 and 6 give the length of the 
finger-of-God and the angular radius assumed to be 
covered by the cluster on the 
sky. Columns 7 and 8 give the number of galaxies in each cluster 
for ORSm and ORSd. 
Figure \ref{Fig8} show the result of this
experiment. The difference in $\xi(s)$ is now clearly reduced (notice
the change in the range in $\xi(s)$ relative to Fig.~\ref{Fig6}).
We thus confirm that $\xi(s)$ depends on selection criteria as long as
small and local volumes, 
containing a significant fraction of low surface brightness/late-types, 
are considered. The presence of high-density regions such as clusters
may also enhance the dependence on selection criteria.
This issue had already been addressed by Zucca \etal (1991),
who found very similar correlation functions for magnitude
and diameter-limited samples in the northern hemisphere, independent
of the value used for $V_{max}$.
Their result, however, was based on volume-limited samples containing
at most 450 galaxies. In addition,
the difference in morphological mix between their
magnitude and diameter-limited samples limited at
$V_{max} = 4000 \kms$ is smaller than in ORSm and
ORSd (see Table~\ref{tab.morph} of this paper 
and Table~2 of Zucca \etal 1991).

\begin{table*}
\caption{Parameters for removed clusters}
\begin{tabular}{|l|c|c|c|r|c|r|r|} \hline
Name &RA & DEC & $V_{mean}$ & $V_{F}$ & radius & ORSm & ORSd \\
     &   &     &(km s$^{-1}$)&(km s$^{-1}$)&(deg.)& & \\ 
Hydra  &158.625 & -27.267 &3450 &  1000 & 3.0  & 40 & 29\\ 
Centaurus&191.525 & -41.033 &3300 &  1000 & 6.0 & 75 & 43\\  
Eridanus& 51.500 & -20.917 &1590 &   400 & 3.0  & 20 & 15\\   
Fornax &53.250 & -35.500 &1350 &   400 & 3.0   &30  & 19\\
Ursa Major& 178.625 & +49.550 &1020 &   300 & 6.0 &37 & 39\\  
Virgo& 187.075 & +12.667 &1080 &  1500 & 6.0  & 148 & 109 \\ \hline
\end{tabular} \label{Tab.NC}
\end{table*}

\subsection{Morphological subsamples}
In this section we investigate the existence of morphological segregation
in ORS. We split each of the five ORS subsamples
compared in section 4.1 into three morphologically 
selected subsamples: early-types ($T \leq 0$), Sab ($1 \leq T \leq 4$) and 
late-types ($T \geq 5$). The number of galaxies in each morphological bin 
for each ORS subsample is given in Table \ref{tab.morph}.
We separately 
fit a selection function to each of these 15 morphological subsamples 
using the method described in Paper II and then merge 
the morphological subsamples together to create the corresponding 
ORSd and ORSm morphological subsamples. We are then left with 6
morphological subsamples, 3 for ORSm and 3 for ORSd.

We thus properly account for the selection effects induced by
the dependence of the luminosity
(or diameter) function on morphology 
and by non-homogeneities in sample selection.
In fact, the luminosity
function is known to vary within the morphological bins defined
above (Binggeli, Sandage \& Tammann 1988, Loveday \etal 1992, Paper II). 
The broad morphological binning chosen in this work reflects
the compromise between this dependence and the 
need to use sizeable samples to reliably derive $\phi({\bf r})$ and $\xi(s)$.
It also has the advantage of
yielding subsamples with roughly similar numbers of 
objects and avoiding systematic differences in morphological
classification likely to exist among the ESO, UGC and ESGC.
In fact, it has been suggested that spirals may have been
misclassified as early-types in UGC due to the lower quality plate material
used for this catalog (Santiago \& Strauss 1992). This effect is likely to be more pronounced at higher redshifts, and in fact some evidence of this
is visible in the type-dependent selection functions used
here.

\begin{figure}
\epsfxsize=8.3cm
\epsfbox{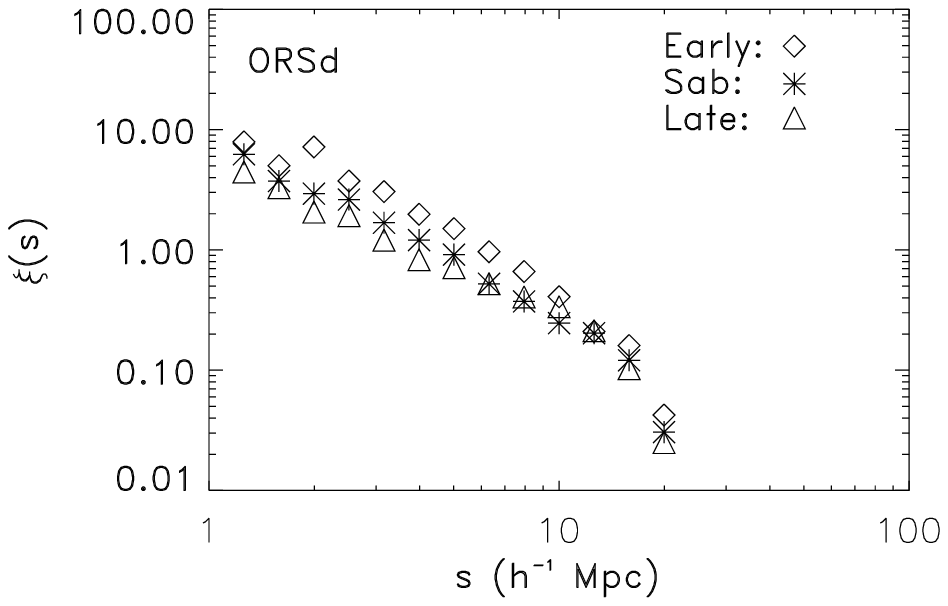}
\caption[]{Correlation functions for three 
morphologically selected subsamples of ORSd. Early types: 
diamonds, Spirals: stars, Late types: triangles.}
\label{fig.ORSdmorph}
\end{figure}

\begin{figure}
\epsfxsize=8.3cm
\epsfbox{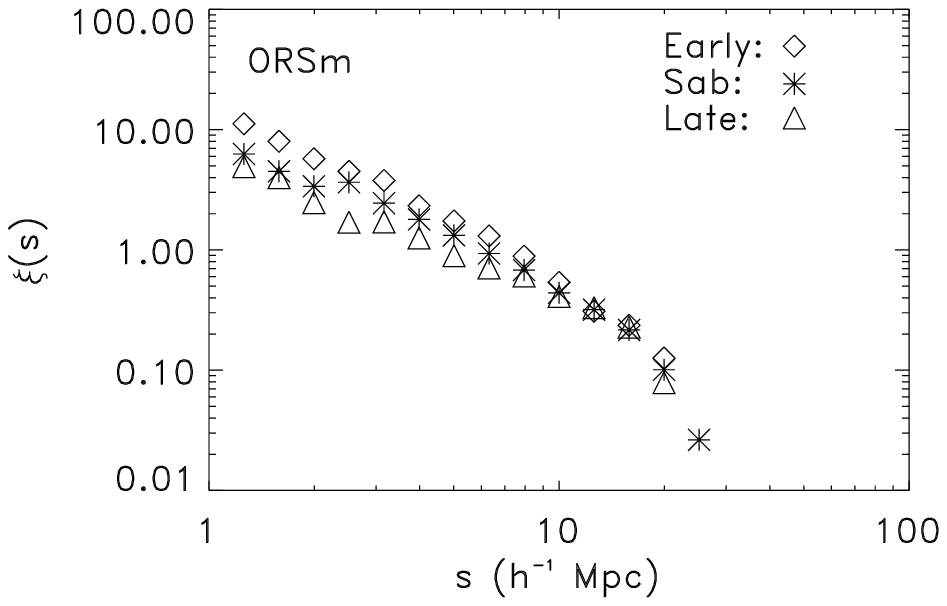}
\caption[]{Same as Fig.~\ref{fig.ORSdmorph}, but for the ORSm.}
\label{fig.ORSmmorph}
\end{figure}

Given the increasing sparseness of late-type galaxies ($T \geq 5$)
beyond $V_{max} = 4000 \kms$, leading to strong shot-noise
problems, we decided to restrict our morphology-clustering investigation
to $V_{max} = 4000 \kms$.
Thus, for each of the six morphological ORS
subsamples we computed $\xi(s)$ using the samples limited at 
$V_{max} = 4000$ km s$^{-1}$. The weighting is again given by the inverse
of the new type-dependent selection functions, with the appropriate
additional factors used when combining the ORS subsamples.
The type correlation functions of ORSd and 
ORSm are shown in Fig.~\ref{fig.ORSdmorph} and \ref{fig.ORSmmorph},
respectively. Results from power-law fits 
are presented in Table \ref{tab.morphfit}.
In each case the range in scales used was
that where $\xi(s)$ was well approximated by a power-law.
 
\begin{table}
\caption{Morphological Subsamples Limited at 4000 km s$^{-1}$ }
\begin{tabular}{|l|c|c|c|c|c|} \hline
Sample & ESGC & ESOd & ESOm & UGCd & UGCm \\
Early    & 141  & 267  & 475  & 223  & 470 \\
Sab      & 142  & 336  & 426  & 353  & 470 \\
Late     & 317  & 542  & 404  & 681  & 617 \\ \hline
\end{tabular} \label{tab.morph}
\end{table}

\begin{table}
\caption{Power-law fits and parameters for the morphological subsamples}
\begin{tabular}{|l|l|l|c|r|r|} \hline
Sample & Type & limit & Range & $\gamma_s$ & $s_{0}$ \\
 & & (h$^{-1}$ Mpc) & ($h^{-1}$ Mpc) &  & \\
    & Early & 40 & 1.3 - 16  & 1.57 & 5.65 \\
ORSd& Sab   & 40 & 1.3 - 16 &  1.53 & 4.24 \\
    & Late  & 40 & 1.3 - 16 &  1.35 & 3.78 \\
 & & & & & \\
    & Early & 40 & 1.3 - 16 &  1.52 & 6.70 \\
ORSm& Sab   & 40 & 1.3 - 16  & 1.32 & 5.63 \\
    & Late & 40 & 1.3 - 16 &  1.17 & 4.71 \\ \hline
\end{tabular} \label{tab.morphfit}
\end{table}

{}From the figures and Table~\ref{tab.morphfit}, we conclude that there is 
strong evidence for morphological segregation out to scales of 
$10 h^{-1}$ Mpc in both 
ORSd and ORSm. We thus confirm the results of many previous works,
e.g., Lahav \& Saslaw
(1992), Santiago \& Strauss (1992), Mo \etal (1992) and Loveday \etal (1995).
The early-type galaxies are more clustered than the Sa/Sb galaxies,
which in turn are more clustered than the late-type galaxies. 
For any given morphological subsample, $\xi(s)$ is 
very similar in both ORSd and ORSm.

\begin{figure}
\epsfxsize=8.3cm
\epsfbox{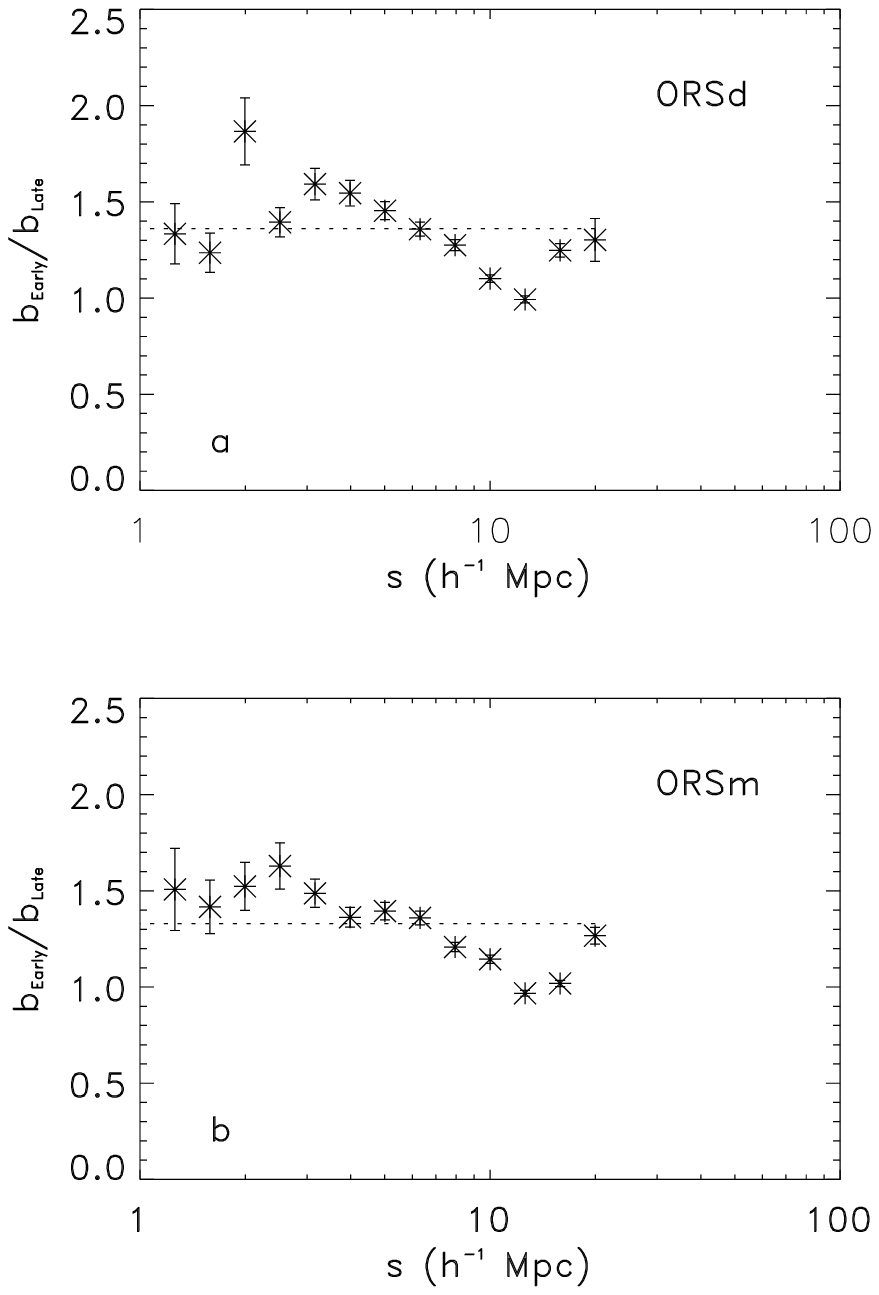}
\caption[]{The relative biasing between early and late type galaxies in the 
ORSd (panel a) and ORSm (panel b). The dotted line is the mean value of 
$b_{Early}/b_{Late}$ in each case.}
\label{fig.relbcomp}
\end{figure}

By comparing the clustering amplitudes of different morphological
subsamples we can infer the corresponding relative bias factor as a 
function of scale using the relation $\xi(s)_{Early} = 
b^{2} \xi(s)_{Late}$.
The scale dependence of $b_{Early}/b_{Late}$ is shown in 
Fig.~\ref{fig.relbcomp} for ORSd and ORSm. In spite of the noise in 
both panels of the figure, the two samples lead to quite similar 
results: $b_{Early}/b_{Late} \sim 1.5$ on small scales ($s$ \ltsima 
$3 h^{-1}$ Mpc), with a declining trend towards larger 
scales, where $b_{Early}/b_{Late} \sim 1$. 
Relative bias factors can also be inferred from the fits made by
Loveday \etal (1995) to their type-dependent $\xi(s)$. A relative 
bias of $\sim 1.4-1.5$ between E/S0 and Spi/Irr is derived from their data
at $s \sim 5 h^{-1}$ Mpc.
Removing the clusters listed in 
Table \ref{Tab.NC} 
makes little difference to the results. A scale dependence is 
also suggested by the UGC and ESO subsamples considered 
individually, indicating that it is not an artifact of the way we 
combined the subsamples or of the weighting scheme used. 
Such a scale dependence is expected from non-linear biasing,
since an enhancement in the number of E/S0s
relative to late-types in small-scale, high density regions such
as clusters contributes to an enhancement in their relative 
pair counts on all scales.

\section{Real-space clustering}\label{sec.xir}
The radial velocities are a combination of the Hubble flow
and peculiar motions. 
The effect of gravitational peculiar motions on small scales 
is to smear out high-density regions such as clusters along 
the radial line-of-sight. On large scales a contraction is 
expected due to infall into high-density regions. This gives 
rise to the well-known $(r_p,\pi)$-diagrams (e.g. 
Fisher \etal 1994, Marzke \etal 1995).
In this section we determine the real-space 
correlation function, $\xi(r)$, in order to assess 
the effect of peculiar motions on the results shown in 
section \ref{sec.compare}.
We essentially follow the prescriptions of Davis \& Peebles (1983).
In brief, we determine the two dimensional 
correlation function $\xi(r_p,\pi)$, where
$r_p$ is the separation perpendicular to the line of sight, and $\pi$ is 
the separation along the line of sight. 
Only pairs separated by less than $50^{o}$ in the sky and $50h^{-1}$ Mpc 
in space were included.
As in $\xi(s)$, we adopted the Hamilton estimator and no
selection effects were applied to the random catalogue.

\begin{figure*}
\epsfxsize=17cm
\epsfbox{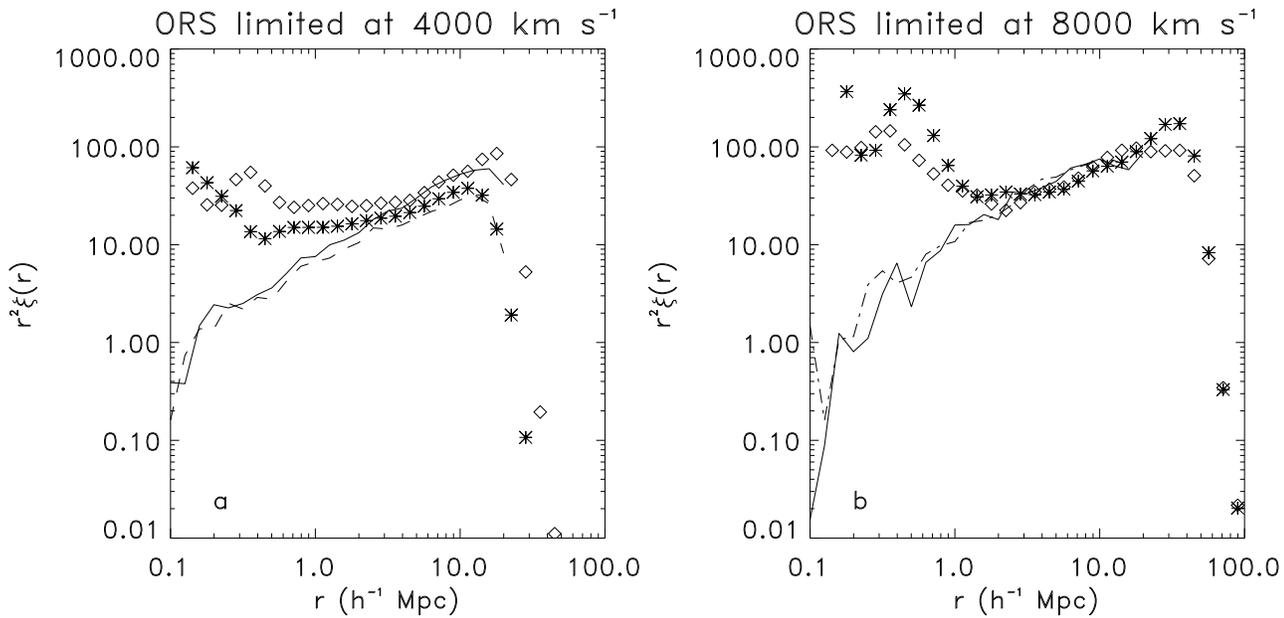}
\caption[]{The real space correlation function $r^2 \xi(r)$ plotted 
against separation $r$ for the ORSd (stars) and ORSm (diamonds) 
limited at $4000 \kms$ (panel a) and $8000 \kms$ (panel b). Shown is also 
the redshift space correlation function, $s^2 \xi(s)$ for ORSd (dashed line) and 
ORSm (solid line).}
\label{fig.wp_xir}
\end{figure*}

Given $\xi(r_p, \pi)$ 
one may obtain the projected function $w(r_p)$ by integrating 
$\xi(r_p,\pi)$ along the $\pi$-direction. 
Once $w(r_p)$ is found one may determine the full $\xi(r)$ by solving 
the Abel integral
\begin{equation}
\xi(r) = - \frac{1}{\pi} \int_{r}^{\infty} { {dw(r_p)} \over {dr_p} } (r_p^2 - 
r^2)^{-1/2} dr_{p}
\end{equation}
numerically.
In Figure \ref{fig.wp_xir} we show our determination of $\xi(r)$ for 
the ORSd and ORSm limited at $4000$ \kms (panel $a$) and $8000$ \kms 
(panel $b$), and compare it to $\xi(s)$. We use $r^2 \xi(r)$ since this
is a less variable quantity with scale of separation.
A number of interesting features are present in this plot.
First, it confirms our results based on redshift space: ORSm is 
more clustered than ORSd for $V_{max} = 4000 \kms$. 
As before, this is not the case for the
samples limited at $V_{max} = 8000 \kms$. Thus, the results shown in
section 4.2 are not an artificiality produced by redshift distortions.
Figure \ref{fig.wp_xir} also shows
that there are significant differences between $\xi(r)$ and $\xi(s)$ in
all samples: $\xi(r)$ is larger than 
$\xi(s)$ on small scales, as expected. 
On large scales, linear perturbation theory gives a proportionality between
$\xi(s)$ and $\xi(r)$, with the proportionality constant a function of
$\beta \equiv {\Omega_0^{0.6}/b}$ (Kaiser 1987). The determination of 
$\beta$ from ORS is beyond the scope of this paper, and
will be presented in a future study.

Finally, we fitted a power law to $\xi(r)$.
We derive $4.9 \leq r_{0} \leq 7.3$ and $1.5 \leq 
\gamma_{r} \leq 1.7$ for the samples shown in
Fig.~\ref{fig.wp_xir}. The best fit values for those cut at $V_{max} =
8000 \kms$ are listed in Table \ref{tab.corr}.

\section{Conclusions}\label{sec.Conclusions} 
Our conclusions are 
\begin{enumerate}

\item  There are significant variations in $\xi(s)$ among the
ORS subsamples limited in velocity at $V_{max} = 4000$ km s$^{-1}$. 
The subsamples limited at $V_{max} = 8000$ 
km s$^{-1}$, on the other hand, yield very similar clustering
patterns, except for the ESGC.
This suggests that ESOd, UGCd, ESOm and UGCm are close 
approximations to a fair samples, at least for the purpose of
estimating $\xi(s)$ for $s$ \ltsima $20 h^{-1}$ Mpc. 
This is further demonstrated by the 
very nice agreement found between $\xi(s)$ as determined here and
those obtained from previous optical surveys covering different 
volumes (Davis \& Peebles 1983, de Lapparent \etal 1988, 
Maurogordato \etal 1992, Loveday \etal 1995).
We thus estimate that a sampling volume 
of $\sim (75 h^{-1} {\rm Mpc})^{3}$ is enough to represent the
low-order clustering properties of the ensemble.

\item We have investigated the clustering properties of 
galaxies as a function of selection criteria and concluded
that significant differences in $\xi(s)$ arise between diameter-limited
and magnitude-limited samples only for small and local volumes,
where diameter-limited samples contain a 
larger fraction of low-luminosity, late-type
galaxies. For larger volumes, both samples are dominated by more
luminous, earlier-type objects, reducing the discrepancy in $\xi(s)$.
These results are largely in agreement
with Zucca \etal (1991), who used smaller samples. 

\item We confirm the existence of morphological segregation out
to scales of $\sim 10 h^{-1}$ Mpc as previously found by
other authors (Lahav \& Saslaw 1992, Mo \etal 1992, 
Dominguez-Tenreiro \etal 1994, Loveday \etal 1995). 
Early-type galaxies ($-5 \leq T \leq 0$)
are more strongly clustered than late-types
($T \geq 5$), whereas Sa/Sb galaxies form an intermediate group.
The difference in clustering amplitudes between E/S0s and Late-types 
suggests that the relative bias factor between these two morphological 
types is weakly dependent on scale. Our results indicate that 
$b_{Early} / b_{Late} \sim 1.5$ on small scales, showing a declining 
trend towards $b_{Early} / b_{Late} \sim 1$ on larger scales 
($s \sim 10 h^{-1}$ Mpc).
Excluding clusters from the morphological samples does not change 
the overall picture; segregation is still present out to
$s \sim 10 h^{-1}$ Mpc and the apparent scale dependence of the
relative bias factor remains. If real, this scale-dependence is 
in contradiction with the popular `linear local biasing' prescription,
in which $b_{Early}$ and $b_{Late}$ are constants.
It would also disagree with the linear limit of the biasing scheme
proposed by Kaiser (1984). Such a scale dependence may be 
explained if we allow for non-linear biasing, e.g. by having a 
higher fraction of early-types formed in high-density regions. 

\item We compute the correlation function 
$\xi(r_p,\pi)$ and from it, find 
$\xi(r)$, the 
real space correlation function. 
This experiment confirms our previous findings in redshift 
space: ORSm galaxies are more clustered than those in ORSd
within $V_{max} = 4000$ \kms but not within $V_{max} = 8000$ \kms.
$\xi(s)$ is also shallower and of smaller amplitude than $\xi(r)$ on
small scales, as expected.
 
\item Our qualitative results are insensitive to the particular kind 
of estimator
used in deriving $\xi(s)$ and to corrections for peculiar velocities.
They are also robust to uncertainties in the selection functions derived
for ORS or to the particular weighting scheme used to compensate for
selection effects or to combine the different subsamples. 

An interesting question is whether the results found
in this work hold when higher order clustering statistics are used.
These latter are also important as means of testing scale-invariance
relations (Bouchet \etal 1993).
The determination of counts in cells and high order correlations is
in progress and will be shown in a forthcoming paper (Hermit \etal 1996, 
in preparation).
\end{enumerate} 

\subsection{Acknowledgements}
This work was carried out at the Institute of Astronomy, Cambridge
University. 
SH also acknowledges support from Julie Marie Vinter Hansens 
Rejselegat, Det Naturvidenskabelige Fakultet and Det 
Internationale Kontor at Copenhagen University. SH would 
like to thank Lars L. Christensen for stimulating discussions.
MAS acknowledges the support of the Alfred P. Sloan Foundation.

\end{document}